\def\year{2022}\relax
\documentclass[letterpaper]{article} 
\usepackage{aaai22}  
\usepackage{times}  
\usepackage{helvet}  
\usepackage{courier}  
\usepackage[hyphens]{url}  
\usepackage{graphicx} 
\usepackage{natbib}  
\usepackage{caption} 
\DeclareCaptionStyle{ruled}{labelfont=normalfont,labelsep=colon,strut=off} 
\usepackage{algorithm}
\usepackage{algorithmic}
\usepackage{subfig}

\usepackage{newfloat}
\usepackage{listings}
\floatstyle{ruled}
\newfloat{listing}{tb}{lst}{}
\floatname{listing}{Listing}

\title{Coordinated through a Web of Images: Analysis of Image-based Influence Operations from China, Iran, Russia, and Venezuela}
\author{
    Lynnette Hui Xian Ng\textsuperscript{\rm 1},
    J.D. Moffitt\textsuperscript{\rm 1}, Kathleen M. Carley\textsuperscript{\rm 1}
}
\affiliations{
    \textsuperscript{\rm 1}School of Computer Science, Carnegie Mellon University\\
    5000 Forbes Ave\\
    Pittsburgh, Pennsylvania 15213\\
    \{huixiann, jdmoffit, kathleen.carley\}@cs.cmu.edu

}

\usepackage{bibentry}

\begin{document}

\maketitle

\begin{abstract}
State-sponsored online influence operations typically consist of coordinated accounts exploiting the online space to influence public opinion. Accounts associated with these operations use images and memes as part of their content generation and dissemination strategy to increase the effectiveness and engagement of the content. In this paper, we present a study of images from the PhoMemes 2022 Challenge originating from the countries China, Iran, Russia, and Venezuela. First, we analyze the coordination of images within and across each country by quantifying image similarity. Then, we construct Image-Image networks and image clusters to identify key themes in the image influence operations. We derive the corresponding Account-Account networks to visualize the interaction between participating accounts within each country. Finally, we interpret the image content and network structure in the broader context of the organization and structure of influence operations in each country. 
\end{abstract}

\section{Introduction}
The global information environment connects the physical, information, and cognitive domains. Engagements in the physical and information domains shape people's world attitudes, beliefs, and world views in the cognitive domain. The rise of the internet age and the development of social media platforms provides state and non-state actors a unique opportunity to exploit the information environment to conduct influence operations \cite{stout2017covert}.

Influence operations can be broadly defined as coordinated, integrated and synchronized efforts to accent communications to affect attitudes and behaviors \cite{larson2009foundations}. Several frameworks have been proposed to characterize influence operations: a process of defining broad objectives then zooming into detailed strategies and deploying negotiation techniques \cite{larson2009foundations}; fusing internet and military concepts in a kill-chain approach \citet{bergh2020understanding}; mapping pathways of information on social media platforms \cite{ng2021does}, and characterizing narrative and network maneuvers \cite{carley2020social}.


Studies have indicated that China, Iran, Russia, and Venezuela have substantial state capacity and resources dedicated to conducting influence operations \cite{bradshaw2019global}. In addition, these countries' coordinated influence operations can be categorized in terms of their content and communication features \cite{alizadeh2020content}. Prior work identified Twitter posts from Russia as spreading propaganda and politically biased information during the 2016 US election \cite{badawy2019characterizing}. There is also strong evidence of automated social media manipulation operations stemming from Iran \cite{bradshaw2017troops}. 

State-sponsored accounts do not rely solely on textual content but also leverage the expressive power of images, e.g., using politically and ideologically-charged memes \cite{rowett2018strategic}. The use of images contributes to increasing message engagement \cite{wang2020understanding}, which can influence online political opinion \cite{doi:10.1080/15358593.2011.653504}. While there is a long stream of work analyzing texts used in influence operations, image analysis in the context of influence operations has been less studied. 

In this paper, we begin filling this gap by analyzing the use of images by state-sponsored accounts in information diffusion. We adopt network-based approaches in analyzing and comparing the coordination structures among the four countries. This approach draws connections between images to indicate their similarity, then further draws connections between accounts when they both coordinate via sharing a similar image. The resultant networks allow visual inspection and comparison of the image and account coordination structure that describes the influence operation of each country. Implementing an unsupervised image comparison and clustering approach provides more flexibility than pre-defined categories in a supervised setting. It is capable of adapting as images and memes evolve during the course of an influence operation. We not only examine the image and account coordination structure through a computational pipeline but also contextualize the observations with the influence operation capabilities of each country as a backdrop.

\textbf{Contributions.}
Using images obtained from identified Twitter inauthentic accounts from four countries, we analyze the image-based influence operations by country and collectively. In this study, we make the following contributions: 
\begin{enumerate}
    \item We develop a systematic methodology for analyzing coordinated image dissemination as part of an influence operation strategy. We analyze the use of images in influence operations in four countries and cross-country image coordination through image similarity networks.
    \item We compare the coordination strategy of accounts in image information dissemination through Account-Account networks across the four countries.
    \item We provide an overview of the image messaging strategy of each country in the broader context using studies on the structure and organization of influence operations in each country.
\end{enumerate}

\section{Related Work}
\textbf{Influence Operations.}
Social media is an effective tool for influencing individuals' opinions \cite{bakshy2011everyone}. The coordinated operation to influence the opinions of large groups of people can threaten the social fabric, as witnessed in the spread of \#StopTheSteal narratives during the 2021 Capitol Riots \cite{DBLP:journals/corr/abs-2109-00945}. Characterizing, understanding, and eventually forecasting how online mediated interactions influence social cohesion and collective decision-making is part of an emerging field of study called social cybersecurity \cite{wang2020defining,carley2020social}. 



The detection of influence operations on social media is no easy task, as these inauthentic state-sponsored accounts put much effort into masking their activity, thus appearing indigenous to their target platform \cite{alizadeh2020content}. Frameworks for detection of coordinated activity based on high levels of common actions have been proposed and studied in the context of high profile events such as the 2016 US Presidential Elections and the 2021 ReOpen America protest campaigns \cite{weber2020s,magelinski2022synchronized}. To handle the evolution of campaigns across time and the differences in influence operations by different groups, supervised machine learning frameworks leveraging content, networks, and user features help identify signals of campaigns \cite{varol2017early}. In order to associate possible sources to the detected influence campaigns, biases in word embedding vectors have been harnessed as a technique to detect target groups \cite{alizadeh2020content}.

Prior research used linguistic analyses to characterize the thematic content and evolution of part-of-speech and stylistic features during the 2016 US Presidential elections \cite{boyd2018characterizing} and detect signatures of state-sponsored influence operations through word embeddings \cite{toney2021automatically}. 

\textbf{Image Analysis.} 
An image speaks a thousand words. Much of the work on image analysis in social media has been in meme identification and comparison \cite{beskow2020evolution}, image sentiment and emotion analysis \cite{wang2015sentiment,10.1007/978-3-030-80387-2_18}, personality analysis through profile pictures \cite{liu2016analyzing}, and profiling natural disasters and epidemics \cite{6389686,nguyen2017damage}. Other image analysis work dwells on breaking down YouTube and TikTok videos into image frames and leveraging their comments to characterize the propagation of themes \cite{hussain2018analyzing,10.1145/3487351.3488276}.

Recent work analyzing themes in image clusters used to incite freedom fighters \cite{10.1007/978-3-030-80387-2_18} and characterizing the content shared by state-sponsored Russian trolls \cite{zannettou2020characterizing} marks a shift in research to leverage image analysis to better understand influence operations. In particular, \citet{zannettou2020characterizing} constructed network graphs representing the similarity of images of inauthentic Russian accounts and observed that these accounts are highly efficient in spreading political images in attempts to sow discord online during dividing events. 

Image similarity assessment techniques typically involved distance-based functions such as the Euclidean distance between vectors containing information about invariant image features \cite{di1999distance}. Image feature signatures have been constructed based on texture, shape, and color of objects\cite{kelly1995query}. Using a combination of image feature values that include grayscale, color, motion, orientation, etc., \citet{manovich2012compare} compared one million digital comic book images and clustered them into themes. We build on this legacy and harness image similarity analysis techniques to characterize image coordination during influence operations. More advanced techniques include simulation to find image variants \cite{morel2009asift}. We build on this rich legacy and harness image similarity metrics to characterize image coordination during influence operations. 

\section{Methodology}
\textbf{Dataset.} 
The images are obtained from the inauthentic accounts dataset from the \texttt{authenticity.training} data from PhoMemes 2022 Challenge\footnote{https://phomemes.github.io}. These images come from state-linked information operations originating from China, Iran, Russia, and Venezuela identified by Twitter Transparency Center \cite{twitterio}. This dataset contains 3801 images from China, Iran, Russia, and Venezuela.

We do not remove repeated images because the presence of repeated images can indicate coordination tactics, i.e., amplification or diffusion of images. However, in the absence of additional metadata about the users or timestamp of the images, we thus focus on the image similarity coordination structure.

\textbf{Ethics.} We worked with the publicly available PhoMemes dataset and did not try to deanonymize account information.

\textbf{Image Feature Extraction.} 
We implement image feature extraction with Python Tensorflow. Each image is resized to a tensor with (244, 244, 3) dimensions as an input to the pre-trained ResNet50 image model \cite{he2016deep}. The image model thus represents the image in terms of a 2048-dimensional vector via 50-layers of convolutional neural networks trained on more than a million images with 1000 object categories. 

\textbf{Image-Image Network.} 
We construct a k-Nearest Neighbors Image-Image similarity network. For each image, we perform a pairwise comparison between the other images in the comparison set by calculating the Euclidean distance between the two image vectors. To ensure flexibility of our method, we set the number of clusters $k$ to be $k=logN$, where $N$ is the number of images of the comparison set, thus adjusting $k$ to the size of the data. A set of $k$ images is said to be similar based on shared image features.

We then constructed an Image-Image network graph $G_I=(V_I, E_I)$ where $V_I$ is the set of nodes $\{V_{I_1}, V_{I_2}...V_{I_n}\}$ representing images and $E_I(V_{I_j}, V_{I_k})$ represents an edge between two image nodes $j$ and $k$ if the images $j$ and $k$ are similar to each other. In short, a node represents an image, and edges represent the extent of similarity between two adjacent images. We lay $G_I$ out with a ForceAtlas2 layout. We did this for images within each country and across all images in the dataset. We interpret networks in terms of (a) clustering coefficient, a measure of the degree to which the nodes in a graph tend to cluster together; and (b) fragmentation, the measure of the disconnectivity of the nodes in the graph.

\textbf{Image Clustering.}
We perform network-based image clustering to identify the key narratives spread through the images. We first segregated the Image-Image network graph using the Louvain clustering algorithm \cite{blondel2008fast}. Then we identify the top six clusters with the most number of images and manually interpret each image cluster.

\textbf{Account-Account Network.}
To represent the coordination of accounts through similar images, we folded the Image-Image network to form an Account-Account To represent the coordination of accounts through similar images, we folded the Image-Image network to form an Account-Account network for each country and all the accounts in the dataset. In this, the Account-Account network graph $G_A=(V_A, E_A)$ where $V_A$ is the set of nodes $\{V_{A_1}, V_{A_2}...V_{A_n}\}$ representing accounts and $E_A(V_{A_j}, V_{A_k})$ represents an edge between two account nodes $j$ and $k$ they have images that are similar to each other. In short, a node represents an account, and edges represent the extent of similar images of two adjacent accounts. We lay $G_A$ out with a ForceAtlas2 layout.

\section{Results}
\begin{figure*}[!tbp]
  \centering
\includegraphics[width=1.0\textwidth]{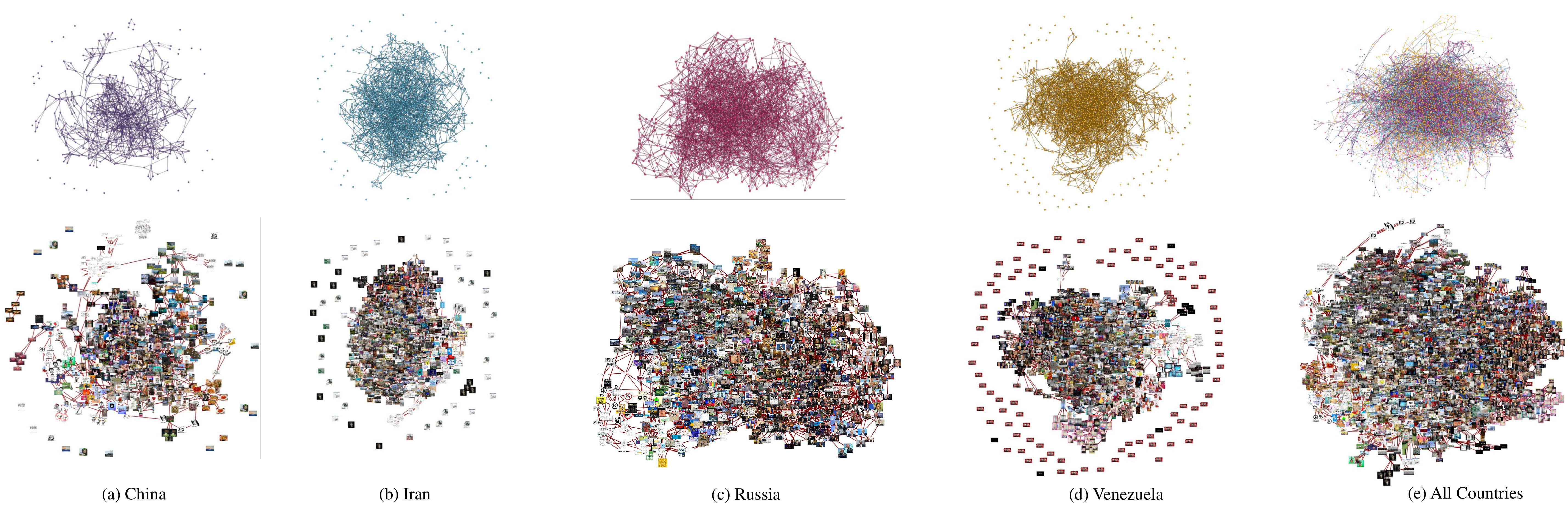}
  \caption{Image-Image Network. Nodes represent images. While the images are small to fit the screen, this figure emphasizes the network structure derived from image similarity metrics. Edge widths represent the similarity between two images. The salient clusters are depicted in Figure \ref{fig:imageclusters}.}
  \label{fig:imageximagenetwork}
\end{figure*}

Our method produces noticeable results in terms of the resulting structure of the image-to-image network, account-to-account network, and the top image clusters. 

\textbf{Image-Image Network.}
We list the network level metrics in Table \ref{table:1}. We use these metrics to reason about the level of coordination in the use of images for influence operations. China has the highest clustering coefficient, indicating a higher coordinated use of images than Iran and Venezuela. All networks but Russia exhibit some fragmentation that may represent noise or attempts not to appear coordinated by state actors. The lack of fragmentation in Russia's network means that every image is similar to at least one other image in the Russian network. 

We visualize the Image-Image networks by country in Figure \ref{fig:imageximagenetwork}. Differences in the clustering coefficient metric can be observed through the network structures: China, Iran, and Venezuela's networks can be visually segmented. In contrast, the network of Russia displays a single centralized cluster. The presence of isolate nodes that line the outer circle of the networks shows the fragmentation of the Image-Image networks and the image themes. 

The image model and similarity measure picks up on general image features such as faces, scenes, and words (memes) in images. This aids us in uncovering visual features used by one or many actors in the same country/event. From the Image-Image network graph of all countries, we observe an interwoven graph, showcasing that images from these four countries generally build on threads from each other. However, some threads are unique to specific countries, represented by the stray network lines sticking out of the core network. 

\begin{figure*}[!tbp]
  \centering
\includegraphics[width=1.0\textwidth]{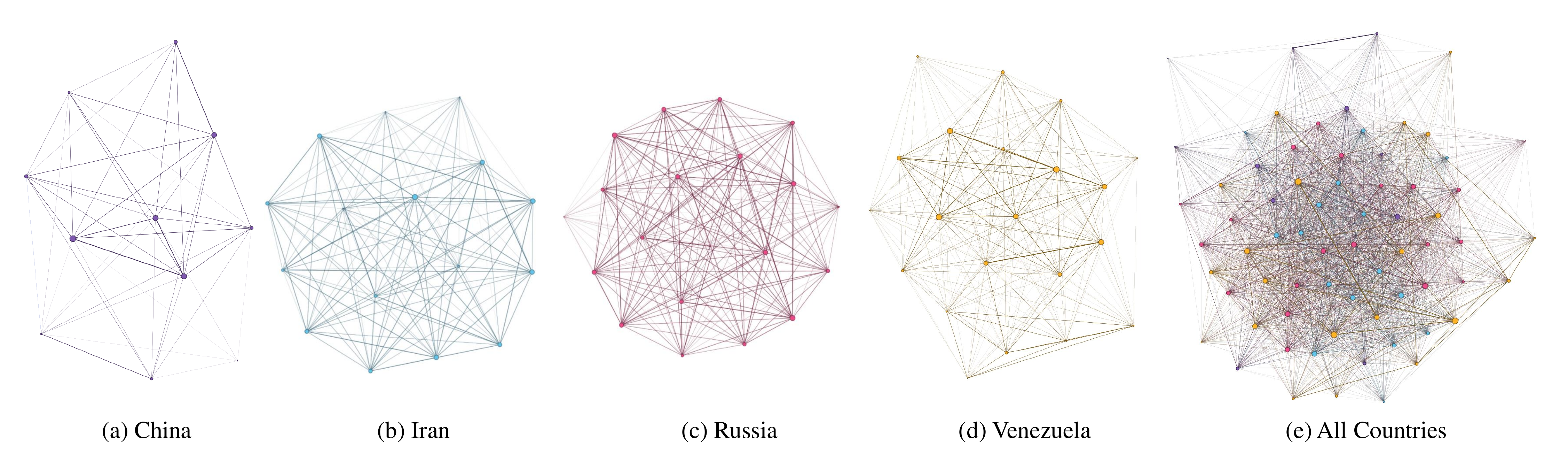}
  \caption{Account-Account Network. Nodes represent inauthentic accounts. Edge widths represent the number of similar images between the two accounts. Adjacent nodes with thicker edges represent a higher number of similar images shared between the two accounts.}
  \label{fig:authorxauthornetwork}
\end{figure*}

\textbf{Account-Account Network.}
In Figure \ref{fig:authorxauthornetwork} we display the Account-Account networks for each country. We find that the networks are densely connected, consistent with the nature of the data, which are accounts identified for coordinated inauthentic behavior by Twitter. Interestingly, we observed varying levels of connectivity between accounts, as evidenced by the edge weights. Accounts with stronger connectivity indicate a larger number of similar image pairs between them. Nodes that are strongly connected may help identify the main coordinators of the influence campaigns. Unfortunately, we do not have the account details in this dataset and are thus unable to identify additional characteristics of the central coordinating accounts.  

\begin{figure*}[!tbp]
  \centering
\includegraphics[width=0.8\textwidth]{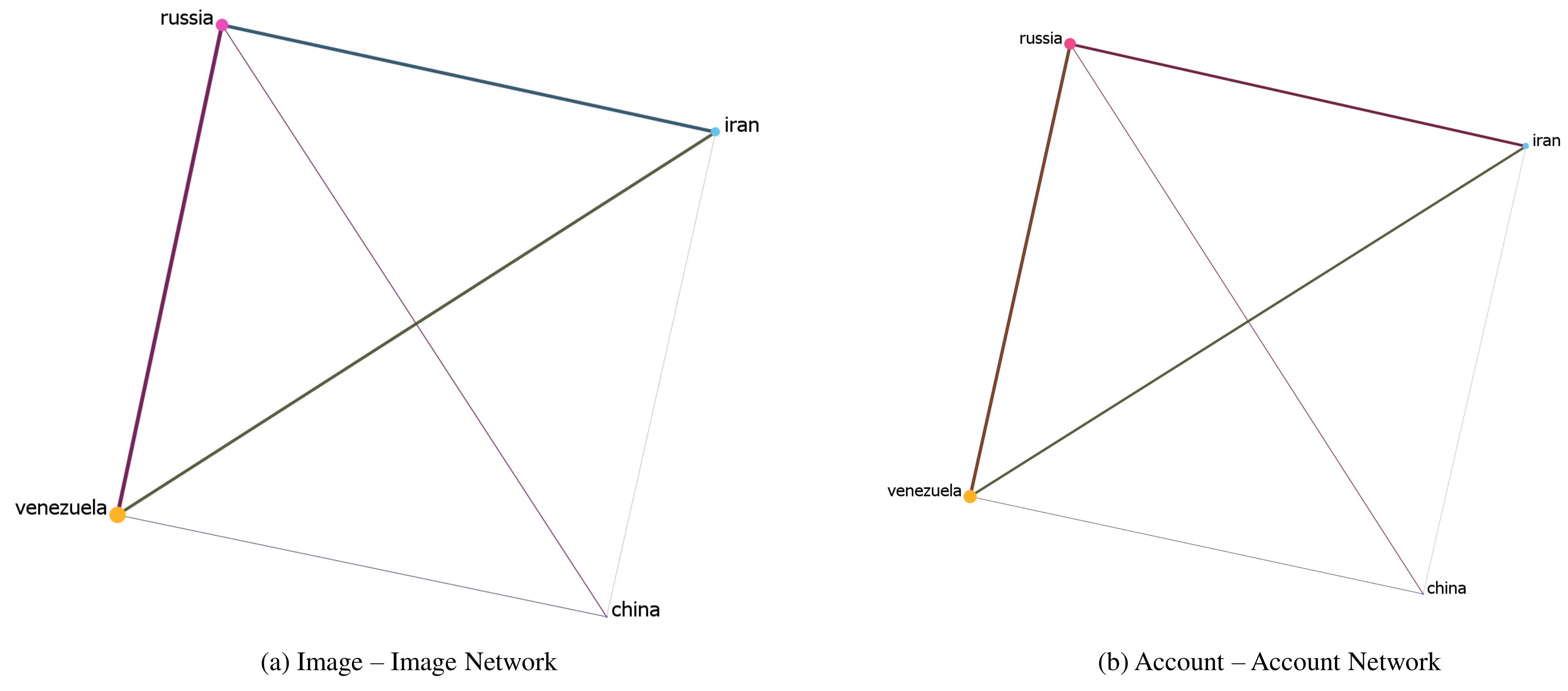}
  \caption{This figure provides the country by country similar image network. In (a), we size nodes by the total number of images, and in (b), we size nodes by the number of unique authors. In (a) and (b), link weights are sized by the number of similar images shared between countries.}
  \label{fig:allauthornetworks}
\end{figure*}

\textbf{Country-Country Network.} We further represent the Image-Image and Account-Account networks for all countries in Figure \ref{fig:allauthornetworks} to showcase the image coordination interaction between all four countries. We observe that Iran, Russia, and Venezuela have strong ties between images and accounts, i.e., many images within these three countries are similar to each other. They have many accounts connected through image similarity.

\begin{table}[h]
\small\sf\centering
\caption{Network level metrics for Image-Image networks}
\label{table:1}
\begin{tabular}{l l l l}
\hline
Country & Images & Clustering Coefficient & Fragmentation \\ 
\hline
China & 598 & \textbf{0.140} & 0.108 \\
Iran & 890 & 0.110 & 0.107  \\
Russia & 1106 & 0.113 & \textbf{0.000}\\
Venezuela & 1207 & 0.112 & 0.149 \\
\hline
\end{tabular}
\end{table}

\begin{figure*}[!tbp]
  \centering
  \subfloat{\includegraphics[width=1\textwidth]{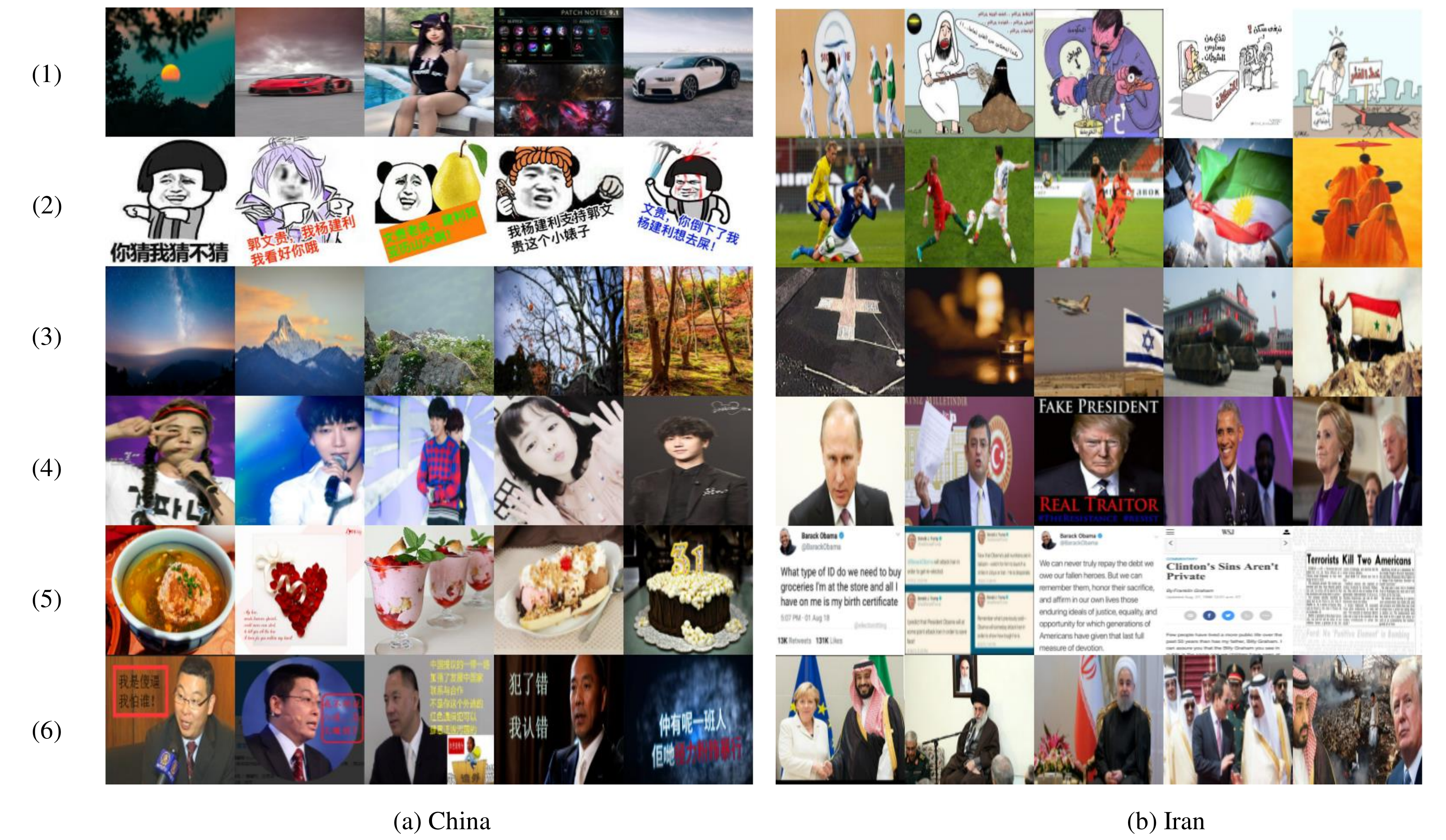}\label{fig:cluster1}}
  \newline
  \subfloat{\includegraphics[width=0.98\textwidth]{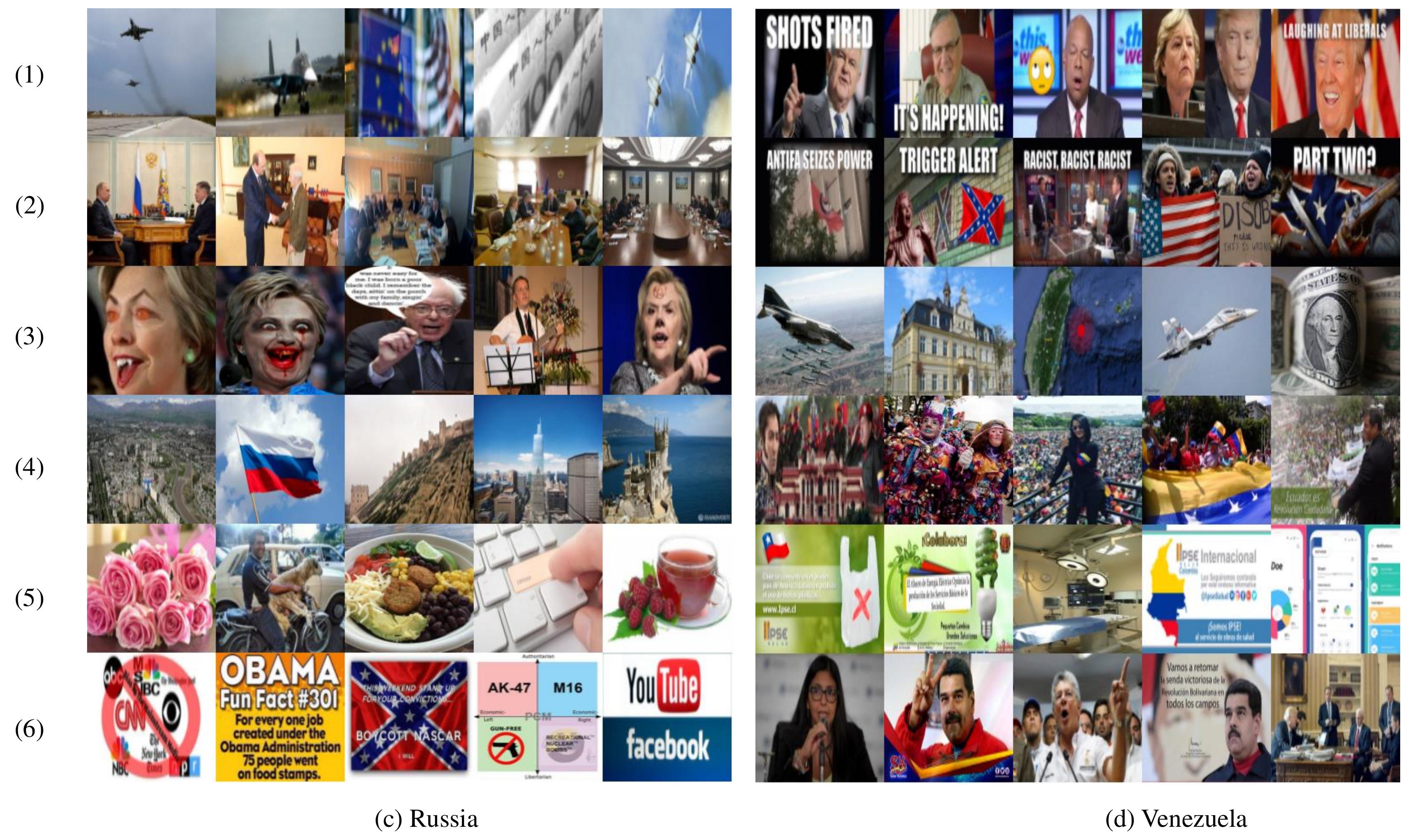}\label{fig:cluster2}}
  \hfill
  \caption{Representative images of the top six image clusters obtained through the Louvain clustering algorithm}
  \label{fig:imageclusters}
\end{figure*}

\textbf{Image Clustering.}
Figure \ref{fig:imageclusters} displays representative images of the top six network clusters obtained through the Louvain clustering algorithm for each country's network. We find that the top clusters reveal distinct and coherent topics for each country. In general, the clusters reveal country-specific themes and influence operations. We also find image clusters of memes -- images with word overlay -- highlighting the use of memes by inauthentic accounts to disseminate information and ideology, probably in a coordinated fashion.



\section{Discussion}
We utilize past forensic evidence and studies on social media behavior, training, and funding estimates as a guide to interpreting our results. We highlight common tactics, account interaction, and communication styles for each country in terms of our image and network analysis results.

\textbf{China} has a high level of coordination and vertical integration from political party to agencies contributing to influence operations \cite{bradshaw2017troops}. There have been estimates that there are over millions of well-trained personnel contributing to influence operations, and the presence of activity-based reward systems \cite{bradshaw2017troops,han2015manufacturing,king2017chinese}. 
Tactics of Chinese influence operations include pro-China messaging, suppressing or censoring messages that counter China's interests, and shifting attention away from essential issues \cite{bradshaw2019global,singer2018likewar}.

Seen through this lens of knowledge about Chinese influence operations, we find three distinct lines of effort in the resulting image clusters for China. First, in row 2, we find memes mocking Guo Wengui, an exiled Chinese billionaire. Second, in row 6, images of the same Chinese billionaire are accompanied by phrases of repentance. Finally, in rows 1, 3, 4, and 5, we find images of lifestyle themes (luxury goods, food, travel, and pop stars). The use of lifestyle type images may be an attempt by influence campaign organizers to leverage popular topic groups of China's rising middle class in an attempt to build a following around its main influence operation \cite{barton2013mapping}, but more importantly, to brand its influence efforts as direct communication with the global public \cite{bund2021finding}. The Image-Image network's structure and high clustering coefficient may correlate to high-level coordination and integration of influence operations.


\textbf{Iran} exhibits a fraction of China's capacity for conducting influence operations, with approximately 20,000 practitioners \cite{bradshaw2017troops}. 
Tactics of Iranian influence operations include pro-government messaging, attacks on the opposition, and suppressing or censoring messages that counter Iran's interests \cite{bradshaw2019global}. Image-based tactics include suppressing political dissidents using political hate speech, vulgar speech, counter-speech, and religion \& societal topics \cite{kargar2019state}.

\textbf{Russia} has a well-funded, well-trained, and highly coordinated influence operation apparatus \cite{bradshaw2017troops}. On top of utilizing similar tactics as discussed with Chinese and Iranian capabilities, the Russian influence operations also focus on driving division and polarization in specific countries or alliances \cite{arif2018acting,thornton2015changing}. 

The Russian Image-Image network is highly connected, with no isolated images, which may further indicate the high-level coordination in the countries' influence operations. Its Account-Account Network is also highly connected and has more edges within the accounts than the other countries' networks. In the resulting image clusters for Russia, we find three distinct lines of efforts. First, evidence of attempts to drive division within the US political landscape in rows 3 and 6 through distorting images of Democratic party candidates and promoting divisive topics in the US media. Second, there are military-based and political-based images in rows 1 and 2, possibly associated with the NATO alliance and its opposition. Finally, we see a line of effort linking lifestyle themes of food and travel, similar to the observations of Chinese influence efforts. 


\textbf{Venezuela} employs multiple bridges of people for influence operations in roughly 500-person teams and utilizes pro-government messaging, attacking the opposition, distraction, and suppression tactics \cite{bradshaw2019global}. There is also evidence that Venezuela is starting to copy Russian approaches to influence operations \cite{lucas2015information}.

The Venezuela Image-Image network features the most isolates, suggesting a good proportion of images are disconnected from the central messaging theme. Many of Venezuela's isolated images are images of "Breaking News," with different news headlines, suggesting one of the Venezuelan strategies is to tweet about breaking news to disguise themselves as a news feed. 

There are varying levels of connectedness between nodes in the Account-Account network, which may result from the organizational structure of the Venezuelan influence operation personnel. 

In terms of image clusters, we see three clear lines of effort. First, in rows 1 and 2, there are political memes. These memes follow a basic template of an image with words written in large white Impact font. In contrast, memes by China have cartoon characters; those by Iran are comic-looking, suggesting that Venezuela's effort in meme generation may be more primitive than the other countries. Second, in row 3, are images of key areas of political tensions like the South China Sea and the United States. Lastly, in rows 4 and 6 are images of protests and rallies. 

\textbf{All Countries}
In Figure \ref{fig:imageclusters} we find that Iran, Russia, and Venezuela have images and memes of U.S. politicians within their top six clusters, while China does not. After folding the Image-Image networks and Account-Account networks, we find strong connections, in terms of shared images, between Iran, Russia, and Venezuela and weaker connections between all three countries and China. Figure \ref{fig:allauthornetworks} displays the folded networks. The high correlation of similar images between Iran, Russia, and Venezuela could indicate coordination between countries or possibly sharing tactics for influence operations. Further research would be needed to substantiate this observation.

\textbf{Limitations and future work.} The primary limitation of this study is the PhoMemes challenge data set itself. The images are studied in isolation from the texts and other meta-data features of the tweets and users, thus losing essential knowledge to understand the intended influence operations better. Additionally, the data may suffer from selection bias and a lack of transparency on how Twitter identified the accounts as part of influence operations. 

The limitations in data provide excellent avenues for future work in studying the use of images in combination with other features in the Tweets. In particular, with tweet and user meta-data, we will construct directed network graphs to identify the diffusion of images across user-to-user all communication networks. Another branch of analysis will entail analyzing influence operations through image networks by characterizing observed changes across time.

\section{Conclusion}
In this study, we characterize the coordinated image dissemination strategy as part of the influence operations of four countries: China, Iran, Russia, and Venezuela. Using image similarity techniques, we construct Image-Image network graphs to identify the coordinated clusters of images. We observe that images from Russia have a single centralized messaging, while images from other countries spout multiple messaging efforts, resulting in isolate nodes in the Image-Image network graphs.
We interpret our network and image cluster observations through the lens of studies about the structure and organization of influence operations in each country, identifying distinct lines of effort in resulting image clusters for each country. By folding the Image-Image network graph, we construct Account-Account network graphs to represent the coordination between accounts through the use of similar images. We observe different levels of connectedness in the network graphs, possibly reflecting the funding and training of the influence operation apparatus of each country.

Grouping images and understanding their links are crucial in large-scale media analytics, aiding deep dives into data subsets for further analysis. Our work provides an unsupervised method to find and characterize groups of accounts and images quickly. This method is a critical first step in a more extensive analysis pipeline for investigating inauthentic coordination activity. We hope that our work provides a preliminary investigation method to aid researchers in understanding the strategic spread of visuals on social media as part of coordinated influence operations.

\section{Acknowledgments}
We would like to thank Jon Storrick from the CASOS, CMU lab for helping us with building a new feature to handle the image visualization on the network visualization software ORA.
The research for this paper was supported in part by the Knight Foundation and the Office of Naval Research grant N000141812106 and by the center for Informed Democracy and Social-cybersecurity (IDeaS) and the center for Computational Analysis of Social and Organizational  
Systems (CASOS) at Carnegie Mellon University. The views and conclusions  are those of the authors and should not be interpreted as representing the official  policies, either expressed or implied, of the Knight Foundation, Office of Naval Research or the US Government. 

\bibliography{aaai22.bib}
\end{document}